\documentclass[11pt]{article}

\usepackage{amssymb,amsmath}
\usepackage{caption,graphicx}
\usepackage[text={6.75in,9.25in},centering]{geometry}
\usepackage[comma,square,sort&compress]{natbib}

\setcaptionmargin{0.25in}

\newtheorem{theorem}{Theorem}
\newtheorem{hypothesis}{Hypothesis}
\newtheorem{lemma}{Lemma}[section]
\newenvironment{proof}[1][.]%
  {\begin{trivlist}\item[]\textbf{Proof#1 }}%
  {\hspace*{\fill}$\rule{.3\baselineskip}{.35\baselineskip}$\end{trivlist}}

\makeatletter\@addtoreset{equation}{section}\makeatother

\newcommand{\N}{\mathbb{N}}								
\newcommand{\R}{\mathbb{R}}								
\newcommand{\Z}{\mathbb{Z}}								

\newcommand{\rmO}{\mathrm{O}}							
\newcommand{\rmo}{\mathrm{o}}							
\newcommand{\rmd}{\mathrm{d}}							
\newcommand{\rme}{\mathrm{e}}							

\def\Fix{\mathop\mathrm{Fix}\nolimits}			
\def\sech{\mathop\mathrm{sech}\nolimits}		
\def\sgn{\mathop\mathrm{sign}\nolimits}			

\begin{document}

\title{Existence of multi-pulses of the regularized short-pulse and Ostrovsky equations}

\author{
Vahagn Manukian\footnote{Present address: Department of Mathematics, University of Kansas, Lawrence, KS 66045, USA  }\\
Department of Mathematics\\
North Carolina State University\\
Raleigh, NC 27695, USA
\and
Nicola Costanzino\\
Department of Mathematics\\
Pennsylvania State University\\
University Park, PA 16802, USA
\and
Christopher K.R.T. Jones\\
Department of Mathematics\\
University of North Carolina\\
Chapel Hill, NC 27599, USA 
\and
Bj\"orn Sandstede\\
Division of Applied Mathematics\\
Brown University\\
Providence, RI 02912, USA
}

\date{\today}
\maketitle

\begin{abstract}
The existence of multi-pulse solutions near orbit-flip bifurcations of a primary single-humped pulse is shown in reversible, conservative, singularly perturbed vector fields. Similar to the non-singular case, the sign of a geometric condition that involves the first integral decides whether multi-pulses exist or not. The proof utilizes a combination of geometric singular perturbation theory and Lyapunov--Schmidt reduction through Lin's method. The motivation for considering orbit flips in singularly perturbed systems comes from the regularized short-pulse equation and the Ostrovsky equation, which both fit into this framework and are shown here to support multi-pulses.
\end{abstract}

\newpage


\section{Introduction}

The motivation for this work comes from the regularized short-pulse equation
\begin{equation}\label{e:r_sort_pulse}
\partial_z \partial_t w + \gamma w + \partial_z^2 w^3 + \beta\partial_z^4 w = 0,
\end{equation}
derived in \cite{SW,CJ} as a model for short pulses in optical fibers, where $w$ is the real component of the envelope of the electric field in the direction transverse to the direction of propagation, $\gamma,\beta\geq0$ are scaled real parameters, obtained from approximating the first order susceptibility tensor for wavelengths in the interval $[1.6\mu\text{m},3.0\mu\text{m}]$, and the Ostrovsky equation
\begin{equation}\label{e:ost1}
\partial_z \partial_t w + \gamma w + \partial_z^2 w^2 + \beta\partial_z^4 w = 0,
\end{equation}
which was derived in \cite{O} for nonlinear waves in the ocean in the small rotation limit and in \cite{OS} to describe oblique magneto-acoustic waves in rotating plasmas. More generally, the equation
\begin{equation}\label{e:rspe1}
\partial_z \partial_t w + \gamma w + \partial_z^2 w^p + \beta\partial_z^4 w = 0,
\qquad p\in\N, \qquad p\geq2,
\end{equation}
which encompasses (\ref{e:r_sort_pulse}) and (\ref{e:ost1}), has been used in \cite{NSC} as a model for nonlinear-wave phenomena in general rotating media.

We are interested in travelling waves of the partial differential equation (\ref{e:rspe1}), which are solutions of the form $w(z,t)=w(z-ct)$, where $c$ is the wave speed. Using the travelling-wave coordinate $y=z-ct$, we find that travelling waves $w(y)$ of (\ref{e:rspe1}) satisfy the ordinary differential equation
\begin{equation}\label{e:profile_ode}
(\beta w_{yy} - cw + w^p)_{yy} + \gamma w = 0.
\end{equation}
It was shown in \cite{CJ} that this equation supports single pulses (that is, homoclinic orbits) in appropriate parameter regions. Our goal is to show that this equation also admits travelling multi-pulses.

The travelling-wave equation (\ref{e:profile_ode}) associated with (\ref{e:rspe1}) is reversible under the operation $y\mapsto-y$ and conservative. We will treat $\beta$, which measures the strength of the regularizing fourth-order derivative term, as a small parameter. This turns the travelling-wave equation into a singularly perturbed problem and makes it amenable to a dynamical-systems analysis. In fact, we will see that the single pulses are in an orbit-flip configuration for $\beta=0$, that is, the single pulses are stronger localized than one would expect. For regular, not singularly perturbed problems, it is known that orbit flips often lead to multi-pulses \cite{S1,S2,SJA,T}. For regular, reversible, conservative systems, it was shown in \cite{S1} that a geometric condition decides whether multi-pulses exist: this condition basically measures whether the conserved quantity allows orbits to follow the primary homoclinic orbit near the origin within the energy surface of the homoclinic orbit. We will encounter a similar condition here for singularly perturbed equations.

A key difference between (\ref{e:r_sort_pulse}) and (\ref{e:ost1}), or more generally between even and odd values of $p$ in (\ref{e:rspe1}), is equivariance under the transformation $w\mapsto-w$, which holds only for odd values of $p$. In particular, the short-pulse equation has two single pulses, which are related by symmetry, while the Ostrovsky equation has only one single-pulse solution. Thus, we may expect that the short-pulse equation supports multi-pulses that follow the two single pulses in an arbitrary order as is the case for orbit-flips in regular reversible systems \cite{SJA}.

To prove our existence and non-existence results, we combine results from geometric singularly perturbation theory with Lyapunov--Schmidt reduction in the form of Lin's method. Lin's method is a tool for constructing multi-pulses that was introduced in \cite{L} and further developed in \cite{S2}. In the travelling-wave context, it has been used, for instance, in \cite{SJA,Y,S2,MS}.

The rest of this paper is organized as follows. In \S\ref{s:2}, we consider equation (\ref{e:profile_ode}) in more depth and use it to motivate the slightly more general setup that we shall consider for our results. Our abstract multi-pulse existence results are stated and proved in \S\ref{s:3} and \S\ref{s:4}, respectively. In \S\ref{s:5}, we apply these results to (\ref{e:profile_ode}).


\section{The profile equations}\label{s:2}

As mentioned in the introduction, we treat the factor $\beta$ in front of the fourth-order term in (\ref{e:profile_ode}) as a singular perturbation parameter. This leads naturally to the scaling
\begin{equation}\label{e:s}
x = \sqrt{\left|\frac{c}{\beta}\right|}y, \qquad
v = |c|^{\frac{1}{1-p}} w,
\end{equation}
which turns (\ref{e:profile_ode}), given by
\begin{equation}\label{e:profile_ode1}
(\beta w_{yy} + w^p - cw)_{yy} + \gamma w =0,
\end{equation}
into the equation
\[
(\sgn(\beta) v_{xx} + v^p - \sgn(c) v)_{xx} + \frac{\gamma|\beta|}{|c|^2} v = 0.
\]
Throughout this paper, we assume that\footnote{When $p$ is odd, our results apply also to the situation $c,\beta,\gamma<0$, upon replacing $v$ by $-v$.}
\begin{equation}\label{e:r1}
c>0, \qquad
\beta, \gamma\geq 0,
\end{equation}
so that the corresponding profile equation is given by
\begin{equation}\label{e:profile1}
(v_{xx} + v^p - v)_{xx} + \epsilon^2 v = 0, \qquad \epsilon^2 = \frac{\gamma\beta}{c^2}.
\end{equation}
When $\epsilon=0$, which corresponds to $\beta\gamma=0$, equation \eqref{e:profile1} supports the pulse solution
\begin{equation}\label{e:pulse1}
q(z) = \left(\frac{p+1}{2}\right)^{\frac{1}{p-1}} \sech^{\frac{2}{p-1}}\left(\frac{p-1}{2}x\right).
\end{equation}
We write \eqref{e:profile1} as the first-order system
\begin{equation}\label{e:fast}
\frac{\rmd}{\rmd x} \left(\begin{array}{c} v_1 \\ v_2 \\ v_3 \\ v_4 \end{array}\right) =
\left(\begin{array}{c} v_2 \\ v_3 + v_1 - v_1^p \\ -\epsilon v_4 \\ \epsilon v_1 \end{array}\right)
\end{equation}
in the coordinates
\[
v_1 = v, \qquad
v_2 = v^\prime, \qquad
v_3 = v^{\prime\prime} + v^p - v, \qquad
-\epsilon v_4 = v_3^\prime.
\]
The first-order system \eqref{e:fast} is Hamiltonian with respect to the skew-symmetric operator
\[
J(\epsilon) =
\left( \begin{array}{cccc} 0 & -1 & 0 & 0 \\ 1 & 0 & 0 & 0  \\
0 & 0 & 0 & -\epsilon \\ 0 & 0 & \epsilon & 0 \\ \end{array} \right)
\]
and the Hamiltonian
\begin{equation}\label{e:h}
H(v_1,v_2,v_3,v_4) :=
\frac{v_1^2}{2} - \frac{v_1^{p+1}}{p+1} + v_1 v_3 - \frac{v_2^2}{2} + \frac{v_4^2}{2}.
\end{equation}
Indeed, using $V=(v_1,v_2,v_3,v_4)^T$, we have
\[
V^\prime = J(\epsilon) H_V(V),
\]
where
\[
H_V(V) = \left( \begin{array}{c} v_3 + v_1 - v_1^p \\ - v_2 \\ v_1 \\ v_4 \end{array} \right).
\]
For $\epsilon=0$, equation (\ref{e:fast}) admits the reversible homoclinic orbit
\[
Q(x) = (q(x),q^\prime(x),0,0)^T
\]
with $q(x)$ from (\ref{e:pulse1}). Note that this solution lies in the fast system of (\ref{e:fast}) and is therefore, by definition, in a singular orbit-flip configuration. It was shown in \cite{CJ} by using geometric singular perturbation theory that this homoclinic orbit persists for $\epsilon\neq0$. Furthermore, it was shown there that, for $\epsilon\neq0$, the persisting homoclinic orbit lies no longer in the strong stable and strong unstable manifolds of the equilibrium $V=0$, but converges to zero as $|x|\to\infty$ with a smaller exponential rate that is proportional to $\epsilon$.

\section{Hypotheses and main result}\label{s:3}

We consider singularly perturbed dynamical systems of the form
\begin{equation}\label{e:sys1}
\frac{\rmd}{\rmd x}
\left(\begin{array}{c} u \\ \tilde{u} \end{array}\right) =
\left(\begin{array}{c} \epsilon f(u,\tilde{u},\epsilon) \\ g(u,\tilde{u},\epsilon) \end{array}\right) = F(U,\epsilon), \qquad 
U = \left(\begin{array}{c} u \\ \tilde{u} \end{array}\right) \in \mathbb{R}^2\times \mathbb{R}^2,
\end{equation}
where $F(U,\epsilon)$ is a smooth nonlinearity, and $(u,\tilde{u})$ denote the slow and fast components of $U$. We now list the assumptions on the system \eqref{e:sys1}.

\begin{hypothesis}\label{h1}
\begin{enumerate}
\item The vector field is reversible, with reverser $R(u_1,u_2,\tilde{u}_1,\tilde{u}_2)=(u_2,u_1,\tilde{u}_2,\tilde{u}_1)$, so that $F(RU,\epsilon)=-RF(U,\epsilon)$ for all $(U,\epsilon)$. Furthermore, we assume that $g(u,\tilde{u},0)=\tilde{g}(u_1+u_2,\tilde{u})$.
\item We assume that there exists a smooth function $H(U,\epsilon):\mathbb{R}^4\times\R\rightarrow\mathbb{R}$ which is invariant under $R$ and satisfies
$\langle H_U(U,\epsilon),F(U,\epsilon) \rangle=0$ for all $(U,\epsilon)\in\mathbb{R}^4\times\mathbb{R}$. We may normalize $H$ so that $H(0,\epsilon)=0$ for all $\epsilon$.
\end{enumerate}
\end{hypothesis}

Hypothesis~\ref{h1}(i) says that the system \eqref{e:sys1} is reversible: if $U(x)$ is a solution of \eqref{e:sys1}, so is $R U(-x)$. Solutions with $U(0)\in\Fix(R)$ satisfy $U(x)=RU(-x)$ for all $x$ and are referred to as reversible. Hypothesis~\ref{h1}(ii) means that (\ref{e:sys1}) is conservative. A particular example of conservative systems are Hamiltonian systems $U^\prime=J(\epsilon)H_U(U,\epsilon)$, where $J(\epsilon):\mathbb{R}^4\rightarrow \mathbb{R}^4$ is a skew-symmetric operator, such that $J(\epsilon)=-J(\epsilon)^\mathrm{T}$. We define
\[
\mathcal{E}_\epsilon := \{U;\;H(U,\epsilon)=0\}
\]
to be the zero level set of $H$. Our next assumption states that $U=0$ is an equilibrium of (\ref{e:sys1}).

\begin{hypothesis} \label{h:fp}
We assume that $U=0$ is an equilibrium of \eqref{e:sys1} for all $\epsilon$ and that the linearization $F_U(0,\epsilon)$ is given by
\[
F_U(0,\epsilon) =
\left( \begin{array}{cccc}
-\epsilon & 0 & 0 & 0 \\
0 & \epsilon & 0 & 0 \\
0 & 0 & -\alpha(\epsilon) & 0 \\
0 & 0 & 0 & \alpha(\epsilon)
\end{array} \right),
\]
where $\alpha(0)>0$. 
\end{hypothesis}

We remark that reversibility makes the spectrum of $F_U(0,\epsilon)$ symmetric with respect to reflections across the imaginary axis. As a consequence of Hypothesis~\ref{h:fp} and geometric singular perturbation theory \cite{J}, the system (\ref{e:sys1}) has a two-dimensional center manifold near $U=0$, and the flow on the center manifold is of the form
\[
u^\prime = \epsilon \left[ \left(\begin{array}{rr} -1 & 0 \\ 0 & 1 \end{array}\right) u + \rm\rmO(|u|^2) \right], \qquad u\in\R^2.
\]
We focus on $\epsilon\geq0$. In this parameter regime, there exists a unique smooth two-dimensional manifold $W^\mathrm{u}(0,\epsilon)$ of (\ref{e:sys1}), which consists of the strong unstable foliation of the one-dimensional unstable manifold within the two-dimensional center manifold. For $\epsilon>0$, this manifold coincides with the usual unstable manifold of $U=0$.

\begin{figure}
\centering
\includegraphics[scale=0.9]{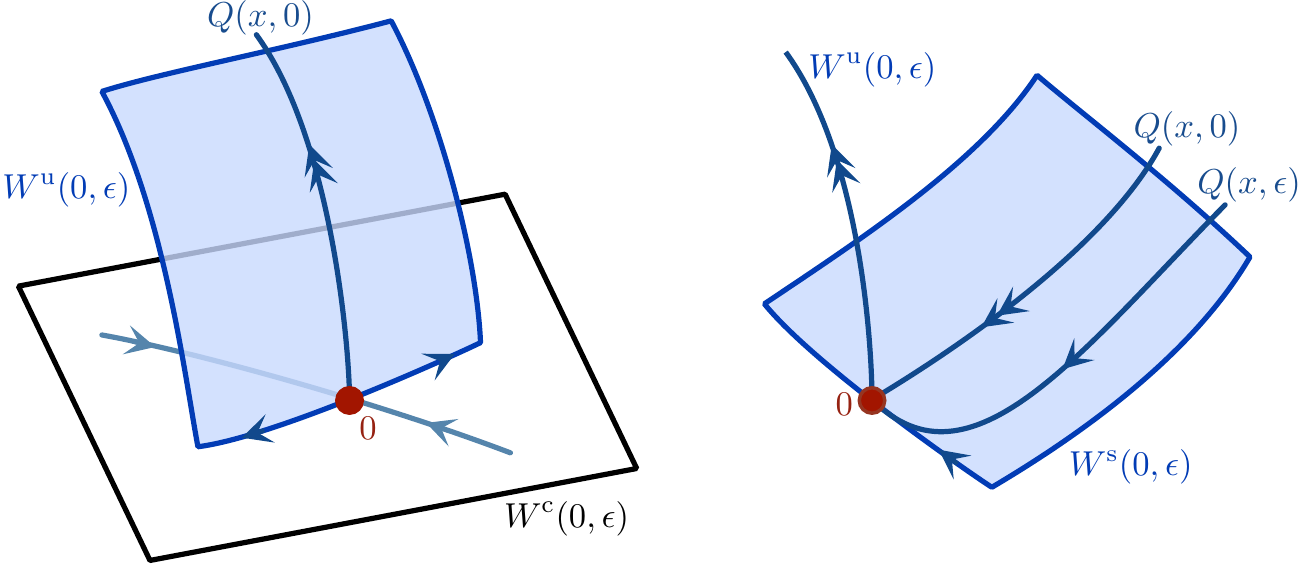}
\caption{The left panel illustrates the geometry of the center manifold and the unstable manifold $W^\mathrm{u}(0,\epsilon)$. The right panel shows how the unfolding of the homoclinic orbit in the stable manifold for $\epsilon\geq0$.}
\label{fig:flip}
\end{figure}

We will assume that the fast system, $\tilde{u}^\prime=g(0,\tilde{u},0)$, has a homoclinic orbit $Q(x,0)$ when $\epsilon=0$. This orbit is automatically transversally constructed with respect to the full system (\ref{e:sys1}), in the sense that $W^\mathrm{u}(0,0)$ and $W^\mathrm{s}(0,0)$ intersect transversally at $Q(0,0)$ inside the level set $\mathcal{E}_0$. Thus, the homoclinic orbit persists for $\epsilon\geq0$. For $\epsilon>0$, it may acquire a slow component $u$, and we shall assume this to be the case; see Figure~\ref{fig:flip} for an illustration. 

\begin{hypothesis} \label{h:flip}
We assume that (\ref{e:sys1}) with $\epsilon=0$ has a reversible homoclinic orbit $Q(x,0)$ to the origin with $H_U(Q(0,0),0)\neq0$. In this case, (\ref{e:sys1}) has a reversible homoclinic orbit $Q(x,\epsilon)$ to the origin for all $\epsilon\geq0$ close to zero, and we assume that
\[
\lim_{x\rightarrow\infty} Q(x,0) \rme^{x} = e_3, \qquad
\lim_{x\rightarrow\infty} Q(x,\epsilon) \rme^{\epsilon x} = \eta^\mathrm{s}(\epsilon) e_1, \qquad
\eta^\mathrm{s}(0) = 0, \qquad
\frac{\rmd\eta^\mathrm{s}}{\rmd\epsilon}(0) \neq 0,
\]
where $e_j$ denote the canonical basis vectors.
\end{hypothesis}

In addition to the assumptions made above, the system (\ref{e:sys1}) may be $\Z_2$-equivariant under the reflection $U\mapsto-U$.

\begin{hypothesis}\label{h2}
We assume that $F(U,\epsilon)$ is odd with respect to $U$ so that $F(-U,\epsilon)=-F(U,\epsilon)$ for all $(U,\epsilon)$.
\end{hypothesis}

\begin{figure}
\centering
\includegraphics[scale=0.9]{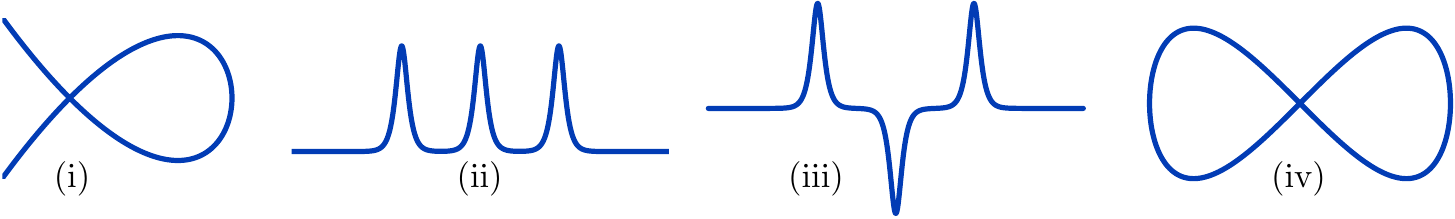}
\caption{Panels~(i) and~(iv) contain phase portraits of homoclinic orbits in reversible systems with~(iv) and without~(i) $\Z_2$-symmetry. Panel~(ii) illustrates the shape of a 3-pulse of up-up type with $\kappa=(1,1,1)$, while panel~(iii) contains an up-down-up 3-pulse with $\kappa=(1,-1,1)$.}
\label{fig:pulses}
\end{figure}

If Hypothesis~\ref{h2} is met, then $Q$ and $-Q$ are both homoclinic orbits, and we may seek $N$-pulses that follow these two orbits in the order given by an arbitrary, but fixed, sequence $\{\kappa_j\}_{j=1,\ldots N}$, where $\kappa_j=\pm1$: the requirement is that the $j$th pulse in the $N$-pulse follows $\kappa_j Q$; see Figure~\ref{fig:pulses} for an illustration. We can now formulate our main result about the existence of multi-pulse solutions.

\begin{theorem}\label{t:exist}
Suppose that Hypotheses~\ref{h1}, \ref{h:fp} and~\ref{h:flip} are satisfied, and define
\begin{equation}\label{e:sgncond}
\sigma := \sgn\left(\langle H_{UU}(0,0)e_3,Re_3\rangle\, \langle H_{UU}(0,0)e_1,Re_1\rangle\right).
\end{equation}
If $\sigma=1$, then $N$-pulses with distances of order $|\ln\epsilon|$ do not exist. If $\sigma=-1$, then, for each $N>0$, there exists an $\epsilon_N>0$ so that \eqref{e:sys1} has an $N$-pulse solution for each $0<\epsilon<\epsilon_N$ that winds $N$ times around the primary pulse solution. The distances $L_j$ between consecutive pulses in this $N$-pulse are given approximately by $L_j\approx-\ln\epsilon$ as $\epsilon\to0$. 

Assume now that Hypothesis~\ref{h2} is also met. If $\sigma=1$, then $N$-pulses with distances of order $|\ln\epsilon|$ do not exist. If $\sigma=-1$, then $2$-pulses of up-up and up-down type exist for $\epsilon>0$; see Figure~\ref{fig:pulses}. Furthermore, for $N>2$, each $N$-pulse either has $\kappa=\pm(1,\ldots,1)$ or else at least one of the distances between consecutive pulses is not of order $|\ln\epsilon|$.
\end{theorem}

The condition $\sigma=-1$ was previously obtained in \cite{S1} for reversible, conservative orbit-flip bifurcations with hyperbolic equilibria. If the product of the scalar products in (\ref{e:sgncond}) is positive, then the energy inside the homoclinic loop in the fast and slow system has the same sign: this may prevent $N$-pulses, which have zero energy, to pass through this area.


\section{Proof of Theorem~\ref{t:exist}}\label{s:4}

Throughout this section, we assume that (\ref{e:sys1}) obeys Hypothesis~\ref{h2}. For a fixed sequence $\{\kappa_j\}_{j=1,\ldots N}$ with $\kappa_j=\pm1$, we will then seek $N$-pulses for which the $j$th pulse in the $N$-pulse follows $\kappa_j Q$. If Hypothesis~\ref{h2} is not met, we simply set $\kappa_j=1$ for all $j$.

\subsection{Fenichel's normal form}

Hypothesis~\ref{h:fp} implies that we can transform the singularly perturbed system (\ref{e:sys1}) near the origin into Fenichel's normal form
\begin{equation}\label{e:fenichel}
\frac{\rmd}{\rmd x}
\left( \begin{array}{c} U^\mathrm{c} \\ U^\mathrm{ss} \\ U^\mathrm{uu} \end{array} \right) =
\left(\begin{array}{c}
\epsilon (A^\mathrm{c}(U^\mathrm{c},\epsilon) U^\mathrm{c} + B(U,\epsilon)[U^\mathrm{ss},U^\mathrm{uu}]) \\
A^\mathrm{ss}(U,\epsilon) U^\mathrm{ss} \\
A^\mathrm{uu}(U,\epsilon) U^\mathrm{uu}
\end{array} \right), \quad
U=\left(\begin{array}{c}U^\mathrm{c}\\ U^\mathrm{ss}\\ U^\mathrm{uu}\end{array}\right) \in \R^2\times\R\times\R,
\end{equation}
where $B(U,\epsilon)$ is a bilinear form for each $(U,\epsilon)$ and
\[
A^\mathrm{c}(0,\epsilon)=\left(\begin{array}{rr} -1 & 0 \\ 0 & 1 \end{array}\right),\qquad
A^\mathrm{ss}(0,\epsilon)=-1, \qquad
A^\mathrm{uu}(0,\epsilon)=1;
\]
see \cite{J} and the references therein. Note that we rescaled $x$ and $\epsilon$ to normalize the fast eigenvalues to be $\pm1$. We remark that the transformation can be chosen so that it is valid in the ball of radius $2$ near the origin and respects reversibility and $\Z_2$ symmetry (when present). Since the slow center manifold, given by $U^\mathrm{ss}=U^\mathrm{uu}=0$, is invariant under (\ref{e:fenichel}), we can straighten out its invariant stable and unstable manifold so that
\begin{equation}\label{e:wc}
A^\mathrm{c}(U^\mathrm{c},\epsilon) U^\mathrm{c} =
\left(\begin{array}{r} A^\mathrm{s}(U^\mathrm{c},\epsilon) U^\mathrm{s} \\
A^\mathrm{u}(U^\mathrm{c},\epsilon) U^\mathrm{u} \end{array}\right), \qquad
U^\mathrm{c}=\left(\begin{array}{c}U^\mathrm{s}\\ U^\mathrm{u}\end{array}\right)\in\R^2,
\end{equation}
where $A^\mathrm{s}(0,\epsilon)=-1$ and $A^\mathrm{u}(0,\epsilon)=1$ for all $\epsilon$. From now on, we will suppress the dependence of $A^j$ and $B$ on the parameter $\epsilon$.

We place two sections, $\Sigma_\mathrm{in}$ and $\Sigma_\mathrm{out}$, at $U^\mathrm{ss}=1$ and $U^\mathrm{uu}=1$, respectively. We are interested in constructing solutions that need time $L_i$ for some large given $L_i$ to pass from $\Sigma_\mathrm{in}$ to $\Sigma_\mathrm{out}$. To find these solutions, we first construct convenient parameterizations of the two-dimensional stable and unstable manifolds of (\ref{e:fenichel}). We parametrize the two-dimensional stable manifold by $Q^+(x,b^\mathrm{s},\epsilon)$ so that $Q^+(L_*,b^\mathrm{s},\epsilon)\in\Sigma_\mathrm{in}$ for all $(b^\mathrm{s},\epsilon)$ for some $L_*>0$, and $Q^+(x,0,\epsilon)=Q(x,\epsilon)$ is the homoclinic orbit. Furthermore, $b^\mathrm{s}$ denotes the $U^\mathrm{s}$-component relative to the homoclinic orbit. Since the stable manifold has $U^\mathrm{u}=U^\mathrm{uu}=0$, we easily find the expansion
\[
Q^+(x,b^\mathrm{s},\epsilon) = \left( [b^\mathrm{s}+\epsilon+\rmO(|b^\mathrm{s}|^2+\epsilon^2)] \rme^{-\epsilon x},0,[1+\rmO(\epsilon)+\rmO(\rme^{-x})]\rme^{L_*-x},0 \right), \qquad x\geq L_*
\]
from (\ref{e:fenichel}) and (\ref{e:wc}): here, we also exploited Hypothesis~\ref{h:flip} and assumed, without loss of generality, that $\rmd\eta^\mathrm{s}/\rmd\epsilon(0)=1$ (this can always be achieved upon replacing $e_j$ by $-e_j$ for $j=1,2$ and rescaling the weak directions). The two-dimensional unstable manifold can be parametrized analogously by $Q^-(x,b,\epsilon)$, where $Q^-(-L_*,b^\mathrm{u},\epsilon)\in\Sigma_\mathrm{out}$, the scalar $b^\mathrm{u}$ lies in the weak unstable direction $U^\mathrm{u}$, and the expansion
\[
Q^-(x,b^\mathrm{u},\epsilon) = \left( 0,[b^\mathrm{u}+\epsilon+\rmO(|b^\mathrm{u}|^2+\epsilon^2)]\rme^{\epsilon x},0,[1+\rmO(\epsilon)+\rmO(\rme^{x})]\rme^{L_*+x} \right), \qquad x\leq-L_*
\]
holds.

Fix a sequence $\{\kappa_i\}_{i=1,\ldots,N}$ of numbers $\kappa_i=\pm1$ that describes how a prospective $N$-homoclinic orbit follows the primary pulse $Q(x,\epsilon)$ and its symmetric counterpart $-Q(x,\epsilon)$. We shall also prescribe the times $2L_i$ that the individual pulses spend near the origin subject to the requirement that
\[
|\epsilon L_i| \leq \rho
\]
for some sufficiently small $0<\rho\ll1$. We seek such $N$-pulses using the parameterization
\begin{equation}\label{e:U}
\begin{array}{lllll}
U_1^-(x)&=&\kappa_1 Q^-(x,b_1^\mathrm{u},\epsilon)&\quad& x\in(-\infty,0),\\
U_i^-(x)&=&\kappa_i [Q^-(x,b_i^\mathrm{u},\epsilon)+V_{i-1}^-(x)]&\quad& x\in(-L_{i-1},0),\\
U_i^+(x)&=&\kappa_i [Q^+(x,b_i^\mathrm{s},\epsilon)+V_i^+(x)]&\quad& x\in(0,L_i),\\
U_N^+(x)&=&\kappa_N Q^+(x,b_N^\mathrm{s},\epsilon)&\quad& x\in(0,\infty)
\end{array}
\end{equation}
and require initially that
\[
U_i^+( L_*) \in \kappa_i \Sigma_\mathrm{in}, \qquad
U_i^-(-L_*) \in \kappa_i \Sigma_\mathrm{out}, \qquad
U_i^+(L_i) = U_{i+1}^-(-L_i)
\]
for all $i$. These conditions mean that $V_i^\pm(x)$ should satisfy
\begin{eqnarray}
& V_i^{+,\mathrm{ss}}( L_*) = 0, \qquad V_i^{-,\mathrm{uu}}(-L_*) = 0 &
\label{e:sigma} \\ \label{e:m}
& \kappa_i [Q^+(L_i,b_i^\mathrm{s},\epsilon)+V_i^+(L_i)] =
\kappa_{i+1} [Q^-(-L_i,b_{i+1}^\mathrm{u},\epsilon)+V_i^-(-L_i)]. &
\end{eqnarray}
In addition, the functions $V_i^\pm(x)$ should satisfy the system
\begin{eqnarray*}
\dot{V}_{i-1}^- & = & F(Q^-(x,b_i^\mathrm{u},\epsilon)+V_{i-1}^-(x),\epsilon) - F(Q^-(x,b_i^\mathrm{u},\epsilon),\epsilon), \qquad x\in(-L_{i-1},0)\\
\dot{V}_i^+ & = & F(Q^+(x,b_i^\mathrm{s},\epsilon)+V_i^+(x),\epsilon) - F(Q^+(x,b_i^\mathrm{s},\epsilon),\epsilon), \qquad\quad x\in (0,L_i),
\end{eqnarray*}
which we write as
\begin{eqnarray}\label{e:veq}
\dot{V}_{i-1}^- & = & F_U(Q(x,b_i^\mathrm{u},\epsilon),\epsilon) V_{i-1}^- + \rmO(|V_{i-1}^-|^2), \qquad x\in(-L_{i-1},0) \\ \nonumber
\dot{V}_i^+ & = & F_U(Q(x,b_i^\mathrm{s},\epsilon),\epsilon) V_i^++\rmO(|V_i^+|^2), \qquad x\in (0,L_i).
\end{eqnarray}
In deriving the above equations, we exploited the $\Z_2$-equivariance of the right-hand side $F(U,\epsilon)$.

We focus on $x\geq0$ and solve the second equation in (\ref{e:veq}) for $x\in(L_*,L_i)$. Using (\ref{e:fenichel}), we obtain the system
\begin{eqnarray}
\dot V_i^{+,\mathrm{c}} & = &
\epsilon \Big(A^\mathrm{c}(Q^{+,\mathrm{c}}(x,b_i^\mathrm{s},\epsilon))V_i^{+,\mathrm{c}}+\rmO(|V_i^{+,c}|^2)+B(V_i^+)[Q^{+,\mathrm{s}}(x,b_i^\mathrm{s},\epsilon)+V_i^{+,\mathrm{ss}},V_i^{+,\mathrm{uu}}]\Big)
\nonumber \\ \label{e:decouple2}
\dot V_i^{+,\mathrm{ss}} & = &
A^\mathrm{ss}(Q^+(x,b_i^\mathrm{s},\epsilon)+V_i^+)(Q^{+,\mathrm{ss}}(x,b_i^\mathrm{s},\epsilon)+V_i^{+,\mathrm{ss}})-A^\mathrm{ss}(Q^+(x,b_i^\mathrm{s},\epsilon))Q^{+,\mathrm{ss}}(x,b_i^\mathrm{s},\epsilon)
\\ \nonumber & = &
A^\mathrm{ss}(Q^+(x,b_i^\mathrm{s},\epsilon))V_i^{+,\mathrm{ss}}+(A_U^\mathrm{ss}(Q^+(x,b_i^\mathrm{s},\epsilon)) V_i^+ + \rmO(|V_i^+|^2))(V_i^{+,\mathrm{ss}}+Q^{+,\mathrm{ss}}(x,b_i^\mathrm{s},\epsilon)))
\\ \nonumber
\dot V_i^{+,\mathrm{uu}} & = &
\Big(A^\mathrm{uu}(Q^+(x,b_i^\mathrm{s},\epsilon))+\rmO(|V^+_i|)\Big)V_i^{+,\mathrm{uu}}.
\end{eqnarray}
We are interested in finding solutions $V_i^+(x)$ of \eqref{e:decouple2} on
$(L_*,L_i)$ so that
\begin{equation}\label{e:bd+}
(V_i^{+,\mathrm{s}},V_i^{+,\mathrm{ss}})(L_*)=0, \qquad
(V_i^{+,\mathrm{u}},V_i^{+,\mathrm{uu}})(L_i)=(a_i^\mathrm{u},a_i^\mathrm{uu})
\end{equation}
for given small $(a_i^\mathrm{u},a_i^\mathrm{uu})$. Proceeding as in \cite{KSS}, we can construct these solutions and find that they obey the expansion
\begin{eqnarray}
V_i^{+,\mathrm{c}}(x) & = & \rme^{\alpha_i^\mathrm{u}(x)} a_i^\mathrm{u} e_2 +\rmO(\rme^{\epsilon(x-L_i)} |a_i^\mathrm{u}|^2 + \epsilon \rme^{-(1-\delta)L_i}|a_i^\mathrm{uu}|)
\nonumber \\ \label{e:v_plus}
V_i^{+,\mathrm{ss}}(x) & = & \rmO(\rme^{-(1-\delta)x} \|V_i^{+,\mathrm{c}}\| +\rme^{-(1-\delta)L_i} |a_i^\mathrm{uu}|) \\ \nonumber
V_i^{+,\mathrm{uu}}(x) & = & \rme^{\alpha_i^\mathrm{uu}(x)}a_i^\mathrm{uu},
\end{eqnarray}
where $\|V_i^{+,\mathrm{c}}\|=\sup_{x\in[L_*,L_i]}|v_i^{+,\mathrm{c}}(x)|$ and
\begin{eqnarray*}
\alpha_i^\mathrm{u}(x) & = & \epsilon \int_{L_i}^x A^\mathrm{c,u}(Q^+(y,b_i^\mathrm{s},\epsilon))\,\rmd y = \epsilon [x-L_i + \rmO(L_i(|b_i^\mathrm{s}|+\epsilon)) ] \\
\alpha_i^\mathrm{uu}(x) & = & \int_{L_i}^x(A^\mathrm{uu}(Q^+(y,b_i^\mathrm{s},\epsilon))+\rmO(|V_i^+(y)|))\,\rmd y = x-L_i +\rmO(L_i(|a^\mathrm{u}|+\epsilon)).
\end{eqnarray*}
We also have
\begin{equation}\label{e:v_c_+}
\|V_i^{+,\mathrm{c}}\| = \rmO(|a_i^\mathrm{u}|+\epsilon \rme^{-(1-\delta)L_i}|a_i^\mathrm{uu}|)
\end{equation}
uniformly in $L_i$. Proceeding in an analogous fashion for $x\leq0$, we can solve the first equation in \eqref{e:decouple2} with boundary data
\begin{equation}\label{e:bd-}
(V_i^{-,\mathrm{u}},V_i^{-,\mathrm{uu}})(-L_*)=0, \qquad
(V_i^{-,\mathrm{s}},V_i^{-,\mathrm{ss}})(-L_i)=(a_i^\mathrm{s},a_i^\mathrm{ss})
\end{equation}
and find that
\begin{eqnarray}
V_i^{-,\mathrm{c}}(x) & = & \rme^{\alpha_i^\mathrm{s}(x)}a_i^\mathrm{s} e_1 +\rmO(\rme^{\epsilon(x+L_i)}|a_i^\mathrm{s}|^2+\epsilon \rme^{-(1-\delta)L_i}|a_i^\mathrm{ss}|)
\nonumber \\ \label{e:v_minus}
V_i^{-,\mathrm{uu}}(x) & = & \rmO(\rme^{(1-\delta)x}\|V_i^{-,\mathrm{c}}\|+\rme^{-(1-\delta)L_i}|a_i^\mathrm{ss}|) \\ \nonumber
V_i^{-,\mathrm{ss}}(x) & = & \rme^{\alpha_i^\mathrm{ss}(x)}a_i^\mathrm{ss},
\end{eqnarray}
where
\begin{eqnarray*}
\alpha_i^\mathrm{s}(x) & = & \epsilon\int_{-L_i}^x A^\mathrm{c,s}(Q^+(y,b_i^\mathrm{u},\epsilon))\,\rmd y = -\epsilon [x+L_i +\rmO(L_i(|b_i^\mathrm{u}|+\epsilon))] \\
\alpha_i^\mathrm{ss}(x) & = & \int_{-L_i}^x(A^\mathrm{ss}(Q^+(y,b_i^\mathrm{s},\epsilon))+\rmO(|V_i^+(y)|))\,\rmd y = - (x+L_i) +\rmO(L_i(|a^\mathrm{u}|+\epsilon)).
\end{eqnarray*}
Again, we have
\begin{equation}\label{e:v_c_-}
\|V_i^{-,\mathrm{c}}\| = \rmO(|a_i^\mathrm{s}|+\epsilon \rme^{-(1-\delta)L_i}|a_i^\mathrm{ss}|)
\end{equation}
uniformly in $L_i$.

\subsection{Construction of Lin orbits}

We now match the solutions $V_i^\pm(x)$ first at $x=L_*$ in $\Sigma_\mathrm{in}$ and then at $x=L_i$ near the origin. To match in the section $\Sigma_\mathrm{in}$, we need to find expansions of the Poincare map that maps $\Sigma_\mathrm{out}$ along the homoclinic orbit $Q(x,\epsilon)$ into $\Sigma_\mathrm{in}$.

\begin{lemma}\label{l:ff}
The Poincare map $\Pi(U,\epsilon)$ from $\Sigma_\mathrm{out}$ to $\Sigma_\mathrm{in}$ satisfies
\begin{eqnarray}\label{e:p}
\lefteqn{ \Pi(U^\mathrm{s},Q(-L_*,\epsilon)+U^\mathrm{u},U^\mathrm{ss};\epsilon) =:
(0,Q(L_*,\epsilon),0) + \tilde{\Pi} } \\ \nonumber & \;=\; &
(0,Q(L_*,\epsilon),0)
+ [1+\rmO(\epsilon)] (U^\mathrm{s},U^\mathrm{u},U^\mathrm{ss})
+ \rmO( (|U^\mathrm{s}|+|U^\mathrm{u}|)|U|,(|U^\mathrm{s}|+|U^\mathrm{u}|)|U,|U^\mathrm{ss}|^2)
\end{eqnarray}
and
\begin{equation}\label{e:sp}
\langle H_U(Q(L_*,\epsilon),\epsilon),\tilde{\Pi}\rangle = \epsilon \sigma U^\mathrm{s} + U^\mathrm{ss} + \rmO(\epsilon|U|(|U^\mathrm{s}|+|U^\mathrm{u}|)+|U^\mathrm{ss}|^2),
\end{equation}
where we use the coordinates $(U^\mathrm{s},U^\mathrm{u},U^\mathrm{ss})$ and $(U^\mathrm{s},U^\mathrm{u},U^\mathrm{uu})$ in $\Sigma_\mathrm{out}$ and $\Sigma_\mathrm{in}$, respectively. 
\end{lemma}

\begin{proof}
The first expansion is really about the form of $\Pi$ when $\epsilon=0$, since the $\epsilon$-dependent terms follow from Taylor expansion. For $\epsilon=0$, equation (\ref{e:sys1}) becomes
\[
u^\prime = 0, \qquad \tilde{u}^\prime = g(u,\tilde{u},0).
\]
Hypotheses~\ref{h1}(i) and~\ref{h:flip} imply that the $\tilde{u}$-equation has a homoclinic orbit for all values of $u$, which implies that $U^\mathrm{ss}=0$ maps into $U^\mathrm{uu}=0$ for $\epsilon=0$. The assertion (\ref{e:p}) follows now from the slow--fast structure and Hypothesis~\ref{h1}(ii). Using the expansion of $Q(x,\epsilon)$, we find that
\[
H_U(Q(L_*,\epsilon),\epsilon)
= \epsilon H_{UU}(0,0) e_1 + H_{UU}(0,0) e_3
= \epsilon \sigma e_2 + e_4
\]
Furthermore, since the gradient of the energy is perpendicular to the stable and unstable manifolds, which are parametrized by $Q(-L_*,b^\mathrm{u},\epsilon)$ and $Q(L_*,b^\mathrm{s},\epsilon)$, we see that there are no contributions of the form $\epsilon b^\mathrm{s}$ or $\epsilon b^\mathrm{u}$ to the scalar product (\ref{e:sp}). On account of the form of the error estimates for $\epsilon=0$, the only nonlinear error terms that can appear in the expansion of the scalar product are as stated in (\ref{e:sp}). 
\end{proof}

Evaluating (\ref{e:v_plus}) and (\ref{e:v_minus}) at $x=\pm L_*$, we obtain
\begin{eqnarray*}
V_i^{+,\mathrm{c}}(L_*) & = & (1+\rmO(\epsilon L_i)) a_i^\mathrm{u} e_2 +\rmO(|a_i^\mathrm{u}|^2 + \epsilon \rme^{-(1-\delta)L_i}|a_i^\mathrm{uu}|) \\
V_i^{+,\mathrm{uu}}(L_*) & = & \rme^{-L_i} a_i^\mathrm{uu} [1+\rmO(L_i(|a_i^\mathrm{u}|+\epsilon))]
\end{eqnarray*}
and
\begin{eqnarray*}
V_{i-1}^{-,\mathrm{c}}(-L_*) & = & (1+\rmO(\epsilon L_{i-1})) a_{i-1}^\mathrm{s} e_1 + \rmO(|a_{i-1}^\mathrm{s}|^2 + \epsilon \rme^{-(1-\delta)L_{i-1}}|a_{i-1}^\mathrm{ss}|) \\
V_{i-1}^{-,\mathrm{ss}}(-L_*) & = & \rme^{-L_{i-1}} a_{i-1}^\mathrm{ss} [1+\rmO(L_{i-1}(|a_{i-1}^\mathrm{s}|+\epsilon))],
\end{eqnarray*}
as well as $V_i^{+,\mathrm{ss}}(L_*)=V_{i-1}^{-,\mathrm{uu}}(-L_*)=0$ due to (\ref{e:bd+}) and (\ref{e:bd-}). To match the corresponding solutions, we transport $Q^-(-L_*,b_{i-1}^\mathrm{u},\epsilon)+V_{i-1}^-(-L_*)$ from $\Sigma_\mathrm{out}$ to $\Sigma_\mathrm{in}$ using the Poincare map $\Pi$ discussed in Lemma~\ref{l:ff}. From (\ref{e:U}), we find that the argument of $\Pi$ is given by
\begin{eqnarray*}
\lefteqn{ Q^-(-L_*,b_{i-1}^\mathrm{u},\epsilon)+V_{i-1}^-(-L_*) } \\ & = &
\left(\begin{array}{c}
(1+\rmO(\epsilon L_{i-1}))a_{i-1}^\mathrm{s}+\rmO(|a_{i-1}^\mathrm{s}|^2+\epsilon\rme^{-(1-\delta)L_{i-1}}|a_{i-1}^\mathrm{ss}|) \\
Q(-L_*,\epsilon)+b_{i-1}^\mathrm{u}+\epsilon+\rmO(|b_{i-1}^\mathrm{u}|^2+\epsilon^2+|a_{i-1}^\mathrm{s}|^2+\epsilon\rme^{-(1-\delta)L_{i-1}}|a_{i-1}^\mathrm{ss}|) \\
\rme^{-L_{i-1}} a_{i-1}^\mathrm{ss}[1+\rmO(L_{i-1}(|a_{i-1}^\mathrm{s}|+\epsilon))]
\end{array}\right).
\end{eqnarray*}
We focus initially on the first two components of $\Pi$ for which Lemma~\ref{l:ff} gives
\begin{eqnarray}\label{e:io}
\lefteqn{ \Pi^\mathrm{s,u}(Q^-(-L_*,b_{i-1}^\mathrm{u},\epsilon)+V_{i-1}^-(-L_*);\epsilon) }
\\ \nonumber & = &
\left(\begin{array}{c} 0 \\ Q(L_*,\epsilon) \end{array}\right) +
\left(\begin{array}{c}
(1+\rmO(\epsilon L_{i-1}))a_{i-1}^\mathrm{s}+\rmO(|a_{i-1}^\mathrm{s}|^2+\epsilon\rme^{-(1-\delta)L_{i-1}}|a_{i-1}^\mathrm{ss}|)
\\ \nonumber
b_{i-1}^\mathrm{u}+\epsilon+\rmO(|b_{i-1}^\mathrm{u}|^2+\epsilon^2+|a_{i-1}^\mathrm{s}|^2+\epsilon\rme^{-(1-\delta)L_{i-1}}|a_{i-1}^\mathrm{ss}|)
\end{array}\right) \\ \nonumber & & 
+ \rmO\left( (|a_{i-1}^\mathrm{s}|+|b_{i-1}^\mathrm{u}|+|\rme^{-L_{i-1}} a_{i-1}^\mathrm{ss}|)(\epsilon+|a_{i-1}^\mathrm{s}|+|b_{i-1}^\mathrm{u}|) \right).
\end{eqnarray}
To simplify the calculations to follow, we anticipate the scalings we shall get: We shall choose
\begin{equation}\label{a:1}
\rho|\ln\epsilon| \leq L_i \leq \frac{\rho}{\epsilon},
\end{equation}
which implies
\begin{equation}\label{a:2}
a_i^\mathrm{s,u}=b_i^\mathrm{s,u}=\rmO(\epsilon), \qquad
\rme^{-L_i}a_i^\mathrm{ss,uu}=\rmO(\epsilon^2).
\end{equation}
Using these estimates, which we shall verify later in the proof, equation (\ref{e:io}) becomes
\[
\Pi^\mathrm{s,u}(Q^-(-L_*,b_{i-1}^\mathrm{u},\epsilon)+V_{i-1}^-(-L_*);\epsilon) =
\left(\begin{array}{c} 0 \\ Q(L_*,\epsilon) \end{array}\right) +
\left(\begin{array}{c}
(1+\rmo(1))a_{i-1}^\mathrm{s}+\rmO(\epsilon^2) \\
b_{i-1}^\mathrm{u}+\rmO(\epsilon^2)
\end{array}\right).
\]
Expanding the first two components of $Q^+(L_*,b_i^\mathrm{s},\epsilon)+V_i^+(L_*)$, and setting them equal to the components of (\ref{e:io}), we arrive at the equations
\begin{eqnarray*}
b_i^\mathrm{s} + \rmO(\epsilon^2) & = & 
(1+\rmo(1))a_{i-1}^\mathrm{s}+\rmO(\epsilon^2) \\
b_{i-1}^\mathrm{u}+\rmO(\epsilon^2) & = &
(1+\rmo(1))a_i^\mathrm{u}+\rmO(\epsilon^2),
\end{eqnarray*}
which we can solve by the implicit function theorem to get
\[
b_i^\mathrm{s} = (1+\rmo(1))a_{i-1}^\mathrm{s}+\rmO(\epsilon^2), \qquad
b_{i-1}^\mathrm{u} = (1+\rmo(1))a_i^\mathrm{u}+\rmO(\epsilon^2).
\]
It remains to match the third components in the $U^\mathrm{ss}$-direction in $\Sigma_\mathrm{in}$. Projecting the difference of $Q^+(L_*,b_i^\mathrm{s},\epsilon)+V_i^+(L_*)$ and $\Pi^\mathrm{s,u}(Q^-(-L_*,b_{i-1}^\mathrm{u},\epsilon)+V_{i-1}^-(-L_*);\epsilon)$ onto $H_U(Q(L_*,\epsilon),\epsilon)$, we obtain
\begin{eqnarray}\label{e:be}
\xi_i & := &
\left\langle H_U(Q(L_*,\epsilon),\epsilon),
\Pi^\mathrm{s,u}(Q^-(-L_*,b_{i-1}^\mathrm{u},\epsilon)+V_{i-1}^-(-L_*);\epsilon) - Q^+(L_*,b_i^\mathrm{s},\epsilon)+V_i^+(L_*) \right\rangle
\\ \nonumber & = &
\epsilon\sigma (1+\rmo(1)) [a_i^\mathrm{u}-a_{i-1}^\mathrm{s}]
+ (1+\rmo(1)) [\rme^{-L_i} a_i^\mathrm{uu} - \rme^{-L_{i-1}} a_{i-1}^\mathrm{ss} ] + \rmO(\epsilon^3).
\end{eqnarray}
We will solve the equations $\xi_i=0$ at the very end of our analysis.

Next, we match the piecewise defined solutions at $x=L_i$. Evaluating \eqref{e:v_plus} at $x=L_i$ and using \eqref{e:v_c_+}, we obtain
\begin{eqnarray*}
V_i^{+,\mathrm{c}}(L_i) & = & a_i^\mathrm{u} e_0^\mathrm{u} + \rmO(\epsilon^2) \\
V_i^{+,\mathrm{ss}}(L_i) & = & \rmO(\epsilon^{2-\delta}) \\
V_i^{+,\mathrm{uu}}(L_i) & = & a_i^\mathrm{uu}.
\end{eqnarray*}
Similarly \eqref{e:v_minus} and  \eqref{e:v_c_-} imply
\begin{eqnarray*}
V_i^{-,\mathrm{c}}(-L_i) & = & a_i^\mathrm{s} e^\mathrm{s}_0 + \rmO(\epsilon^2) \\
V_i^{-,\mathrm{ss}}(-L_i) & = & a_i^\mathrm{ss} \\
V_i^{-,\mathrm{uu}}(-L_i) & = & \rmO(\epsilon^{2-\delta}).
\end{eqnarray*}
The matching condition (\ref{e:m}),
\[
\kappa_i [Q^+(L_i,b_i^\mathrm{s},\epsilon)+V_i^+(L_i)] =
\kappa_{i+1} [Q^-(-L_i,b_{i+1}^\mathrm{u},\epsilon)+V_i^-(-L_i)],
\]
then becomes
\begin{eqnarray*}
\kappa_i
\left(\begin{array}{c}
(\epsilon + b_i^\mathrm{s} + \rmO(\epsilon^2)) \rme^{-\epsilon L_i} \\
a_i^\mathrm{u} + \rmO(\epsilon^2) \\
\rme^{-L_i} + \rmO(\epsilon^{2-\delta}) \\
a_i^\mathrm{uu}
\end{array}\right)
= \kappa_{i+1}
\left(\begin{array}{c}
a_i^\mathrm{s} + \rmO(\epsilon^2) \\
(\epsilon + b_{i+1}^\mathrm{u} + \rmO(\epsilon^2)) \rme^{-\epsilon L_i} \\
a_i^\mathrm{ss} \\
\rme^{-L_i} + \rmO(\epsilon^{2-\delta})
\end{array}\right).
\end{eqnarray*}
We find
\[
a_i^\mathrm{ss} = \kappa_i\kappa_{i+1} \rme^{-L_i} + \rmO(\epsilon^{2-\delta}), \qquad
a_i^\mathrm{uu} = \kappa_i\kappa_{i+1} \rme^{-L_i} + \rmO(\epsilon^{2-\delta})
\]
and
\[
a_i^\mathrm{s}=
\kappa_i\kappa_{i+1} (\epsilon + a_{i-1}^\mathrm{s} + \rmO(\epsilon^2))(1+\rmo(1)), \qquad
a_i^\mathrm{u}=
\kappa_i\kappa_{i+1} (\epsilon + a_{i+1}^\mathrm{u} + \rmO(\epsilon^2))(1+\rmo(1)).
\]
We can solve the preceding system of equations to get
\[
a_i^\mathrm{s} =
\epsilon\kappa_{i+1} \left(\rmo(1)+\sum_{j=1}^i \kappa_j\right),\qquad
a_i^\mathrm{u} =
\epsilon\kappa_{i} \left(\rmo(1)+\sum_{j=i+1}^N \kappa_j\right),\qquad
i=1,\ldots N-1.
\]
Note that this validates the first relation in (\ref{a:2}) that we used above. We now substitute these expressions into the remaining bifurcation equations (\ref{e:be}), which become
\begin{eqnarray}\label{e:ben}
\xi_i & = &
\epsilon\sigma (1+\rmo(1)) [a_i^\mathrm{u}-a_{i-1}^\mathrm{s}]
+ (1+\rmo(1)) [\rme^{-L_i} a_i^\mathrm{uu} - \rme^{-L_{i-1}} a_{i-1}^\mathrm{ss}] + \rmO(\epsilon^3)
\\ \nonumber & = &
\epsilon^2\kappa_i\sigma (1+\rmo(1))
\left[ \sum_{j=i+1}^N \kappa_j - \sum_{j=1}^{i-1} \kappa_j \right]
+ (1+\rmo(1)) \kappa_i \left[\kappa_{i+1} \rme^{-2L_i} - \kappa_{i-1} \rme^{-2L_{i-1}}\right]
+ \rmO(\epsilon^3)
\end{eqnarray}
for $i=1,\ldots,N$, where $L_0=L_N=\infty$.

\subsection{Bifurcation equations}

On account of \cite[Lemma~3.2]{SJA}, it suffices to solve (\ref{e:ben}), given by
\[
\epsilon^2\sigma (1+\rmo(1))
\left[ \sum_{j=i+1}^N \kappa_j - \sum_{j=1}^{i-1} \kappa_j \right]
+ (1+\rmo(1))
\left[\kappa_{i+1} \rme^{-2L_i} - \kappa_{i-1} \rme^{-2L_{i-1}}\right]
+ \rmO(\epsilon^3) = 0,
\]
where $i=1,\ldots,N-1$, since $\xi_N$ then vanishes automatically due to the presence of the conserved quantity $H$ that we assumed to exist. Define $a_i>0$ via
\[
\epsilon^2 a_i = \rme^{-2L_i},
\]
and observe that this yields (\ref{a:1}) and the remaining second estimate in (\ref{a:2}) that we used to derive the bifurcation equations. We obtain
\begin{equation}\label{be:1}
\sigma (1+\rmo(1)) 
\left[ \sum_{j=i+1}^N \kappa_j - \sum_{j=1}^{i-1} \kappa_j \right]
+ (1+\rmo(1)) \left[\kappa_{i+1} a_i - \kappa_{i-1} a_{i-1} \right]
+ \rmO(\epsilon) = 0,
\end{equation}
where $i=1,\ldots,N-1$. Setting $\epsilon=0$, we arrive at the system
\begin{equation}\label{be:2}
\kappa_{i+1} a_i - \kappa_{i-1} a_{i-1} =
\sigma \left[ \sum_{j=1}^{i-1} \kappa_j - \sum_{j=i+1}^N \kappa_j \right]
\end{equation}
or, written out in detail, at
\begin{eqnarray*}
\kappa_2 a_1 & = &
-\sigma (\kappa_2+\ldots+\kappa_N) \\
-\kappa_{i-1}a_{i-1}+\kappa_{i+1}a_i & = &
-\sigma (\kappa_{i+1}+\ldots +\kappa_N) + \sigma(\kappa_{i-1}+\ldots+\kappa_1) \\
-\kappa_{N-2}a_{N-2}+\kappa_N a_{N-1}  & = &
-\sigma \kappa_N
+ \sigma(\kappa_{N-2}+\ldots+\kappa_1)
\end{eqnarray*}
for $i=2,\ldots,N-2$. The left-hand side is linear and invertible in $a=(a_1,\ldots,a_{N-1})$, as it corresponds to a lower triangular matrix with strictly positive entries $\kappa_{i+1}$ on the diagonal. Thus, if we can find a positive solution $a_i>0$ to (\ref{be:2}) for a given sequence $\kappa_i$, then we can solve the full equation (\ref{be:1}) using the implicit function theorem.

First, we shall look for $N$-pulses of up-up type and therefore set $\kappa_i=1$ for all $i$, so that (\ref{be:2}) becomes
\[
a_i - a_{i-1} = -\sigma (N+1-2i), \qquad i=1,\ldots,N-1.
\]
The equation for $i=1$ is $a_1=-\sigma (N-1)$, which has a positive solution only when $\sigma=-1$. Thus, $N$-pulses of up-up type can exist only for $\sigma=-1$. Hence, we take $\sigma=-1$ and therefore need to solve
\[
a_i - a_{i-1} = N+1-2i, \qquad i=1,\ldots,N-1.
\]
This system has the positive solution $a_i=i(N-i)>0$ with $i=1,\ldots,N-1$, since
\[
a_i - a_{i-1} = i(N-i) - (i-1)(N-(i-1)) = i(N-i) - (i-1)(N-i+1) = N+1-2i.
\]
The preceding discussion therefore shows that there is a $N$-pulse of up-up type for each $\epsilon>0$ and that there are no other $N$-pulses of up-up type whose distances satisfy (\ref{a:1}).

Next, we investigate $2$-pulses, when Hypothesis~\ref{h2} is met. In this case we need to solve the single equation
\[
\kappa_2 a_1 = -\sigma\kappa_2
\]
which has the solution $a_1=-\sigma$, independently of $\kappa_2$. Thus, for $\sigma=-1$, $2$-pulses of both up-up ($\kappa_1=\kappa_2$) and up-down ($\kappa_1=-\kappa_2$) type exist.

It remains to consider the case $N>2$ for arbitrary sequences $\{\kappa_i\}$.

\begin{lemma}
Without loss of generality, let $\kappa_2=1$, and assume that $\kappa_i\in\{\pm1\}$ for $i=1,\ldots,N$, where $N>2$. If (\ref{be:2}) has a solution $a=\{a_i\}_{i=1,\ldots,N-1}$ with $a_i>0$ for all $i$, then necessarily $\sigma=-1$ and $\kappa_i=1$ for all $i$.
\end{lemma}

Note that this completes the proof of Theorem~\ref{t:exist} due to the arguments presented above, upon taking the assumption (\ref{a:1}) about the distances between consecutive pulses into account.

\begin{proof}
Let
\begin{equation}\label{e:b}
b_i = \sigma \left[ \sum_{j=1}^{i-1} \kappa_j - \sum_{j=i+1}^N \kappa_j \right],
\qquad i=1,\ldots,N-1.
\end{equation}
Using induction, we find that
\[
b_i = b_1 +\sigma \left[ \kappa_1 + \kappa_i + 2\sum_{j=2}^{i-1} \kappa_j \right],
\qquad i\geq2.
\]
Similarly, an induction argument shows that the solution of (\ref{be:2}) is given by
\begin{equation}\label{e:a}
a_i = \kappa_i \kappa_{i+1} \sum_{j=1}^i \kappa_j b_j,
\qquad i=1,\ldots,N-1.
\end{equation}
Using the normalization $\kappa_2=1$, we find that $a_1=\kappa_2 b_1=b_1>0$, hence $b_1\geq1$. We now proceed again by induction and make the following induction statement at the $i$th step:
\[
\sum_{j=1}^i \kappa_j b_j = i[b_1+\sigma(i-1)], \qquad\qquad
b_1 \geq \left\{\begin{array}{lcl} 1 & & \text{when }\sigma=1 \\
i & & \text{when }\sigma=-1, \end{array}\right. \qquad\qquad
\kappa_j = 1 \text{ for } j\leq i+1.
\]
Using (\ref{e:a}) and the assumption that $a_j>0$ for all $j$, it is not difficult to check that the statement is true for $i=2$ and to carry out the induction step from $i$ to $i+1$, and we therefore omit the details. Thus, we find that $\kappa_i=1$ for all $i$, which implies $b_1=-\sigma(N-1)$. When used in combination with $b_1\geq1$, these statements prove the lemma.
\end{proof}


\section{Application to the generalized Ostrovsky equation}\label{s:5}

We now return to the generalized Ostrovsky equation (\ref{e:rspe1})
\begin{equation}\label{e:goe}
\partial_z \partial_t w + \gamma w + \partial_z^2 q^p + \beta\partial_z^4 q = 0,
\qquad p\in\N, \qquad p\geq2.
\end{equation}
As we had seen in \S\ref{s:2}, travelling waves of the form $w(z,t)=w(z-ct)$ satisfy the differential equation (\ref{e:profile_ode1})
\[
(\beta v_{yy} + v^p - cv)_{yy} +\gamma v = 0,
\]
where $y=z-ct$ is the travelling coordinate. The scaling (\ref{e:s}) and the condition (\ref{e:r1}), namely $c>0$ and $\beta,\gamma\geq0$, allowed us to write this equation as (\ref{e:profile1}),
\[
(v_{xx} + v^p - v)_{xx} + \epsilon^2 v = 0, \qquad \epsilon^2=\frac{\gamma\beta}{c^2},
\]
which supports the pulse
\begin{equation}\label{e:q}
q(x) = \left(\frac{p+1}{2}\right)^{\frac{1}{p-1}} \sech^{\frac{2}{p-1}}\left(\frac{p-1}{2}x\right)
\end{equation}
when $\epsilon=0$. Equivalently, we can write this equation as the first-order system (\ref{e:fast})
\begin{equation}\label{e:ode}
\frac{\rmd}{\rmd x} \left(\begin{array}{c} v_1 \\ v_2 \\ v_3 \\ v_4 \end{array}\right) =
\left(\begin{array}{c} v_2 \\ v_3 + v_1 - v_1^p \\ -\epsilon v_4 \\ \epsilon v_1 \end{array}\right)
\end{equation}
Equation (\ref{e:ode}) is reversible with respect to the reverser $R(v_1,v_2,v_3,v_4)=(v_1,-v_2,v_3,-v_4)$ and conservative with respect to the first integral (\ref{e:h}),
\[
H(v_1,v_2,v_3,v_4) =
\frac{v_1^2}{2} - \frac{v_1^{p+1}}{p+1} + v_1 v_3 - \frac{v_2^2}{2} + \frac{v_4^2}{2},
\]
and it is now easy to check that Hypothesis~\ref{h1} is met. It was shown in \cite{CJ} that Hypotheses~\ref{h:fp} and~\ref{h:flip} are met for the pulse given in (\ref{e:q}).
Furthermore, we have
\[
\mathrm{D}^2H(0) =
\left(\begin{array}{rrrr} 1 & 0  &  1  &  0 \\ 0 & -1 & 0 & 0 \\
1 & 0 & 0 & 0 \\ 0 & 0 & 0 & 1 \end{array}\right),
\]
and, using that the slow and fast stable eigenvectors of the linearization of (\ref{e:ode}) about the origin are given by
\[
e_\mathrm{s}  = \left(1,0,-1,-1\right)^T, \qquad
e_\mathrm{ss} = \left(1,-1,0,0\right)^T,
\]
we find that
\[
\sgn \left( \langle \mathrm{D}^2H(0) e_\mathrm{s},R e_\mathrm{s}\rangle\;
\langle \mathrm{D}^2H(0) e_\mathrm{ss},R e_\mathrm{ss}\rangle \right) = -1.
\]
Finally, (\ref{e:ode}) is equivariant with respect to $V\mapsto-V$ if and only if $p$ is odd. We can therefore apply Theorem~\ref{t:exist} to get the following result:

\begin{theorem}
Fix $\gamma,c>0$. For each $N\geq2$, there exists a $\beta_N>0$ such that the generalized Ostrovsky equation (\ref{e:goe}) has an $N$-pulse $w_N(z-ct)$ of up-up type for each $\beta$ with $0<\beta<\beta_N$. The distances between consecutive pulses in the $N$-pulse $w_N(z-ct)$ are of the order $-\beta\ln\beta$. If $p$ is odd, $-w_N(z-ct)$ is also an $N$-pulse, and $2$-pulses of both up-up and up-down type exist for $0<\beta<\beta_2$.
\end{theorem}


\end{document}